\documentclass[12pt]{iopart}

\usepackage{graphicx} 
\usepackage{bbm}
\usepackage{amssymb}                           
\begin{document}

\title{Social and strategic imitation: the way to consensus}

\author{Daniele Vilone$^{1,*}$, Jos\'e J.\ Ramasco$^1$, Angel S\'anchez$^{2,3}$, Maxi San Miguel$^1$}
\address{$^1$  IFISC (CSIC-UIB), Campus Universitat de les Illes Balears,  07122
Palma de Mallorca, Spain}
\address{$^2$ Grupo Interdisciplinar de Sistemas Complejos (GISC),
Departamento de Matem{\'a}ticas, Universidad Carlos III de Madrid,
28911 Legan{\'e}s, Spain}
\address{$^3$ Instituto de Biocomputaci{\'o}n y F{\'\i}sica de Sistemas
Complejos (BIFI), Universidad de Zaragoza, 50018 Zaragoza, Spain}

\ 

\address{$^*$ Correspondence and requests for materials should be
addressed to D.V.~(email: dvilone@ifisc.uib-csic.es).}

\maketitle

\begin{abstract}
Humans do not always make rational choices, a fact that experimental economics is putting on solid grounds. The social context plays an important role in determining our actions, and often we imitate friends or acquaintances without any strategic consideration. We explore here the interplay between strategic and social imitative behaviors in a coordination problem on a social network. We observe that for interactions in 1D and 2D lattices any amount of social imitation prevents the freezing of the network in domains with different conventions, thus leading to global consensus. For interactions in complex networks, the interplay of social and strategic imitation also drives the system towards global consensus while neither dynamics alone does. We find an optimum value for the combination of imitative behaviors to reach consensus in a minimum time, and two different dynamical regimes to approach it: exponential when social imitation predominates, and power-law when strategic considerations dominate.
\end{abstract}

%{\color{red} En la relectura hay que ver que no se confunda las estrategias del juego (+1, -1) con el concepto de imitacion estrategica. ...Quizas al referirse al juego hablar solo de coordination choices o options)}

\ 

When facing a choice, it is often the case that people do not make the optimal decision \cite{roth:1995,fehr:2000,camerer:2003,gintis:2009}. Within the framework of economics, the explanation for these deviations has been advanced in terms of social preferences: Agents are not solely motivated by material self-interest but also care positively or negatively for the material benefits of their counterparts \cite{fehr:2002}. Models with alternative individual utility functions have also been put forward in an attempt to incorporate the experimental evidence \cite{charness:1999,fehr:1999}. Yet another line of thought has resorted to evolutionary arguments to explain non-selfish behavior \cite{henrich:2001,gintis:2003,sanchez:2005}. In spite of this progress towards a quantitative understanding of human behavior, a number of important questions remain unanswered. One such question relates to the effect of the interplay between strategic or economic views and social factors to explain people choices in a game. 

A very similar question has been also considered from a different perspective. 
From a sociological viewpoint, there have been many attempts to uncover the mechanisms underlying the adoption by people of new technologies, the acceptance and spreading of rumors or in general of new information~\cite{granovetter:1978,Holley:1975,klemm:2003,castellano:2009,vilone:2004,Suchecki:2005}. For example, a basic model in this context, proposed by Granovetter \cite{granovetter:1978}, assumes that a certain amount of social pressure is necessary for a person to adopt a new concept. The pressure in this model, as in other opinion models, is quantified as the number of contacts that have already adopted the concept.

Indeed, imitation is an important mechanism through which the social environment may influence strategic decisions. Imitation has been related to bounded rationality or to a lack of information that compels agents to copy the strategies of others~\cite{schlang:1998}.  Several proposals have been advanced to describe imitation. For example, it has been proposed that players might imitate one of their counterparts with  a probability  proportional to the difference in the agents'~benefits~\cite{helbing:1992}. Interestingly, this assumption leads the game to a dynamics equivalent to that of the so-called replicator equation~\cite{schlang:1998,helbing:1992,nowak:2006}. Another method to include imitation in a game is the so-called unconditional imitation. This means that after each round of the game an agent copies the strategy of her coplayer with highest payoff as long as it is higher than her own. However, it is clear that imitation may not be perfect for many reasons; therefore, in previous works, the effect of mixing unconditional imitation with other dynamics such as random strategy selection has been studied in the context of binary choices~\cite{roca:2009b,traulsen:2009}. Random decisions help the system to explore different actions and such exploration may in fact lead to higher cooperation in public goods or prisoner's dilemma games. Similarly, random changes in models of cultural evolution (cultural drift) allow to escape from frozen states of cultural polarization~\cite{klemm:2003}. 

Our aim in this paper is to go beyond pure randomness and consider, instead of pure random noise, the interplay of two possible imitation dynamics: One is strategic and modeled by the  unconditional imitation driven by the game payoff; the other is of social nature and is inspired by the voter model. In the voter model~\cite{Holley:1975,castellano:2009,vilone:2004,Suchecki:2005}, an agent simply copies the opinion of a randomly selected counterpart. This mechanism favors the spreading of a majority option, which in our case will be related to the action taken. The opinion update rule incorporates thus the effect of the social pressure regardless of the payoffs obtained in the previous game round. The social component introduces new features in the dynamics of the game that may lead to different final configurations of the system~\cite{gargiulo:2012}. It is also important to stress that, in contrast to random strategy selection (noise), the voter model generates a correlated state of opinion in the social network. 

We address the above issue by introducing a model in which social and strategic considerations drive individual behaviors and study what occurs depending on the frequency of every type of decision procedure. To be specific, we choose as our strategic problem a coordination game (CG)~\cite{cooper:1999,blume:2002,vega:1997}. Coordination is relevant in many daily actions, from the choice of side on which to drive to the decision on which technology to rely on through opting for a particular phone provider. In what follows, we will focus on the pure coordination game setup, in which the binary choice takes place among equivalent options. This will allow us to achieve a better understanding of the interplay of socially motivated and strategic decisions. As we will see below, even in this simpler setting such interplay will lead to non-trivial, unexpected, results.

\section*{Results}       

\subsection*{Model description}

When playing a pure coordination game, the desired goal of all players is to make the same choice as their counterparts. Even if the choice is not the optimal, it is better to coordinate on a sub-optimal action than to do the opposite of what the other players do. In this work we consider the simplest version of such a coordination game: When two individuals play, they choose between two possible actions, obtaining a payoff of $1$ if they choose the same action and $0$ otherwise. Players are located in the nodes of a social network, which represents their social context. In other words, the people with whom they have to interact and, eventually, to try to achieve coordination. Every player plays a coordination game with each of her neighbors in the network, subject to the constraint that the chosen action is the same for all those games. Of course, when the game is played in a heterogeneous  network, the best decision to make depends on the number of opponents choosing every action. Still, the optimal situation both individually and globally is that all players make the same decision.

To this game-theoretical setup, we have to add the dynamics, namely, the manner in which players update their choices in time. Actually, this is the key point we are analyzing, so let us describe this aspect in detail. Firstly, strategic decisions, i.e., those aimed at improving the payoff players obtain from the game, take
place by means of theunconditional imitation (UI) rule as described above~\cite{nowak:1992}: After every round, players imitate the action of their best performing neighbor, provided that such neighbor receives a larger payoff than the player herself. Subsequently, payoffs are set to zero, a new instance of the game is played, and so on. The final configuration of a population evolving in this manner depends not only on the evolution rule, but also on the topology. For example, in a well mixed population (described by a complete network, implying that every player interacts directly with every other one) perfect coordination is reached in one time step (consensus); on the other hand, in one- and two-dimensional regular lattices the dynamics leads to disordered (non-coordinated) frozen configurations, while on complex networks the precise
topology of the system can either enhance or hinder the reaching of complete coordination~\cite{roca:2009,roca:2010}.

The updating procedure we have described in the preceding paragraph is of a strategic nature, driven by the goal of improving one's payoff. To this behavior, we incorporate another of social nature: An imitative, non-strategic dynamics solely driven by social considerations in which players imitate others without considering how this will affect their payoffs. Such an update rule can be well described by the voter model (VM)\cite{clifford:1973}: At every round of the game, a neighboring agent is picked up at random and the player imitates her choice. It is important to keep in mind that the voter dynamics in $d=1,2$ regular lattices orders the system. Spatial domains of each of the two possible coordinated states grow in time (unbounded growth in the infinite size limit). This is described by dynamical laws for the average density of active links $n_A(t)$. The active links are defined as those  connecting agents with different choices. On the contrary, in the well mixed case, on lattices with
dimension $d\geq3$ and also in complex networks of high effective dimensionality there is no continuous growth of domains of coordination. The voter dynamics leads to a dynamical metastable disordered state with a constant value for $n_A(t)$ in which there is short range coexistence of the two equivalent coordination options. The system remains in this long lived metastable state up to a time proportional to the system size, in which finite size fluctuations take the system to a fully coordinated state ($n_A(t)$=0). Note that this implies that the system remains disordered in the infinite size limit\cite{frachebourg:1996,castellano:2003,Suchecki:2005,suchecki:2005b,vazquez:2008}.

In order to make progress towards our goal of understanding the relative role and importance of the two kinds of dynamics, we  consider a system of $N$ agents on a graph. Each agent can interact only with her nearest neighbors in the network. At each elementary time step, we pick up an agent $i$ at random that plays the game with her neighbors, as they also do with their own neighbors. Once the game is over and a payoff value is assigned the agent $i$ updates her choice.  She does so ``socially'', i.e., according to the
voter dynamics with probability $q$, and strategically by unconditional imitation with probability $1-q$. 

\subsection*{Simulations}

We have carried out a thorough simulation program considering different forms for the social networks on which players interact. We begin the summary of our results by reporting on the two simplest situations, low dimensional regular lattices and complete networks, which will allow us to develop our first intuitions of the mechanisms controlling the model behavior.

\begin{figure}
\begin{center}
\includegraphics[angle=0,width=10cm,clip]{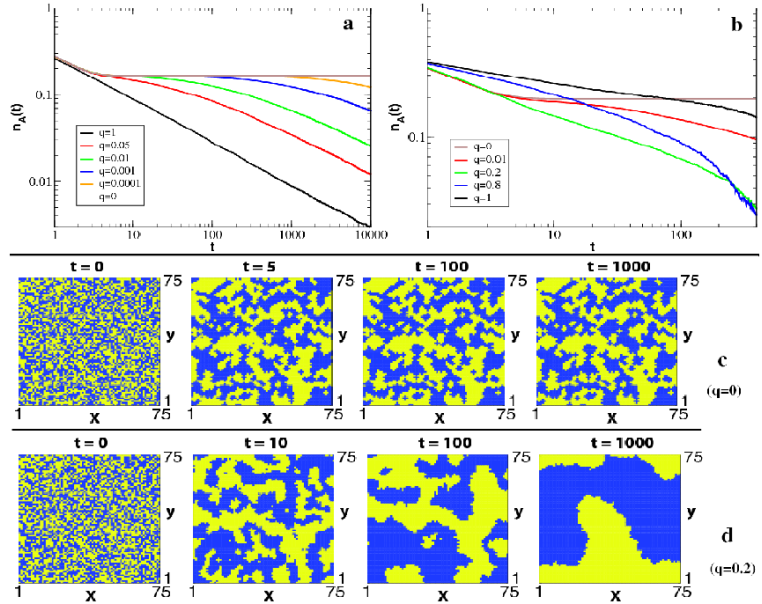}
\caption{Decay of the active links with time: a, for a unidimensional system ($N=1000$) with different values of $q$;
b,  for a bidimensional system ($N= 30\times 30 = 900$). c and d, diagrams showing the system evolution in two dimensional lattices with $N = 75\times 75 = 5625$ and two values of $q$: c for $q = 0$ and d for $q = 0.2$. Yellow and blue represent the two possible choices. The initial conditions for the actions are the same in both panels and are selected at random.
\label{fig1}}
\end{center}
\end{figure}

{\em One- and two-dimensional lattices}. In the limit of pure voter dynamics, $q = 1$, the system orders with $n_A(t)$ approaching zero as $t^{-1/2}$ in $d=1$ or $(\ln t)^{-1}$ in $d=2$, while
in the limit $q = 0$ (pure unconditional imitation) a frozen disordered state is the ultimate fate of the system.
The temporal evolution of $n_A(t)$ for arbitrary values of $q$ is displayed in Figure~\ref{fig1} for lattices of one and two dimensions. A first aspect to notice from pannels \ref{fig1}a and \ref{fig1}b is that  consensus is always reached as long as $q > 0$. As can be observed in panels c and d of Figure~\ref{fig1}, the mechanism responsible for the consensus is the nucleation of the initial domains containing agents making a homogeneous choice. Competition between neighboring domains with opposite choices then takes place until eventually the fluctuations and finite system size lead to a symmetry breaking and to the selection of a single option as dominant. The introduction of the voter dynamics is the main responsible for the symmetry breaking, irrespective of how low $q$ is  in so far as it is non-zero.  A similar picture applies to the one-dimensional case.

\begin{figure}
\begin{center}
\includegraphics[angle=0,width=10cm,clip]{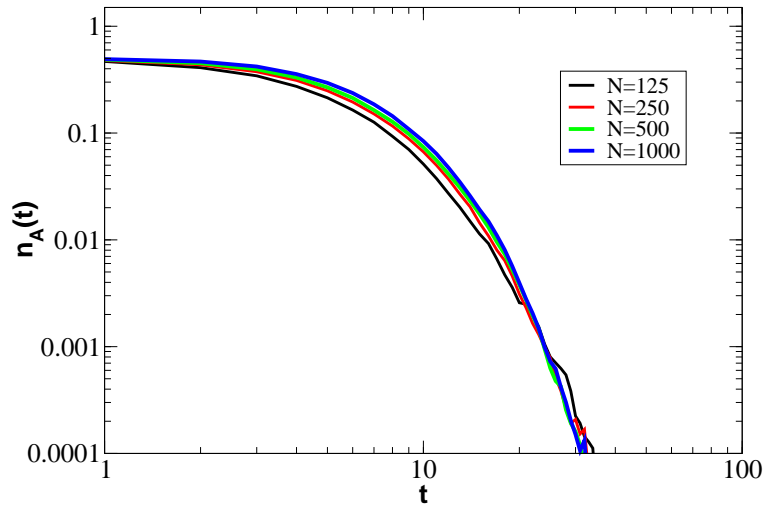}
\caption{Active bond decay for a system in an Erdos-Renyi network ($\langle k\rangle=9$) for $q=0.7$ and different system sizes. The results are averaged over $100$ realizations.
 \label{fig2} }
\end{center}
\end{figure}

As a second social network, we now turn to {\em complete graphs} in which each agent interacts with every other agent. The results in this topology correspond to the mean-field limit of the model. In this topology, the two limits of $q$ in the model behave in an opposite manner. The two options survive in a dynamic state if the update is done with the voter dynamics alone $q=1$, while the consensus is reached in one step for the case of unconditional imitation,  $q=0$. Note that this is similar to the case of low dimensional lattices, although inverting the outcome of the $q$ limits. If the value of $q < 1$, the symmetry between both choices is broken and one option eventually becomes dominant. In this case, it is the presence of unconditional imitation the factor that helps to break this symmetry and leads to the ordering the system.

The two previous network topologies may be regarded as benchmarks since they have very particular properties that facilitate their numerical and analytical treatment. However, they cannot be taken as valid models for more realistic social interactions. Most of the empirical evidence points to a topological organization of social networks as {\em sparse complex graphs}. To take this into account, we now proceed with the study of the game on networks generated with the Erd\"os-R\'enyi (ER) and the Molloy-Reed algorithms~\cite{molloy:1995}. The main difference between both type of random networks is the heterogenity of the number of nodes' connections (degree). The distribution of nodes' degrees in ER graphs is Poissonian, while in the Molloy-Reed networks it decays as a power-law with an exponent $\beta$.

\begin{figure}
\begin{center}
\includegraphics[angle=0,width=10cm,clip]{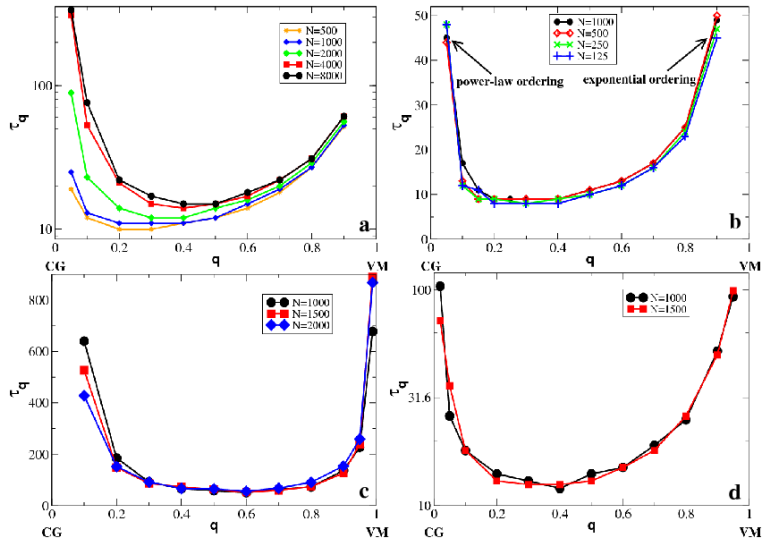}
\caption{The characteristic time for ordering $\tau$ as a function of $q$. a: for Erd\"os-R\'enyi network with $\langle k\rangle=9$ and different system sizes. b: same as in a but $\langle k\rangle=14$.
c: scale-free network with exponent $\beta=3.5$, and
d: scale-free network with $\beta=2.5$. Note the logarithmic scale of the vertical axis in panel a and d. 
\label{fig3}}
\end{center}
\end{figure}

The sparsity of links and the small world effect bring an interesting feature to our mixed model. In contrast to low dimensional lattices and complete graphs, the dynamics does not order the system in any of the two limits of $q$.  The voter dynamics alone ($q = 1$) displays a long lived dynamical state, while the game with only unconditional imitation ($q = 0$) falls into a non-coordinated frozen state. Surprisingly, our simulation results show that the combination of both imitation dynamics (strategic + social) changes the final outcome of the system. When both types of imitation are combined,
the system orders reaching a fully coordinated state in a time which does not scale with system size. We have developed some analytical approaches to understand how this occurs that will be discussed below. In a nutshell, the combination of voter stochasticity and unconditional imitation, which drives the system towards the creation of local majorities, is responsible for the selection of one of the two choices and its spreading. As a representative example of the time evolution of the system, Fig.\ \ref{fig2} shows  the time evolution of $n_A(t)$ for a dynamics with an intermediate value of $q$ and for different system sizes. It can be  seen how the system orders relatively fast and that increasing the system size beyond a certain value of $N$ does not substantially change the picture.

\begin{figure}
\begin{center}
\includegraphics[angle=0,width=10cm,clip]{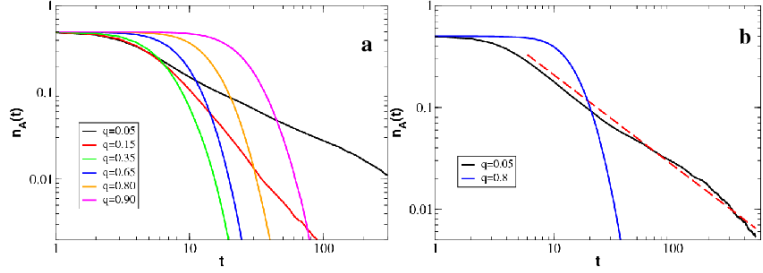}
\caption{Active link decays for a system in an Erd\"os-R\'enyi network with different values of $q$. In a, for $\langle k\rangle=9$ and $N=8000$, while in b
with $\langle k\rangle=9$ and $N=16000$.
\label{fig4}}
\end{center}
\end{figure}

As a practical way to measure the velocity with which the system reaches full coordination, we define the characteristic time  $\tau$ as the time in which the density of  active links falls below $n_A(t) = 10^{-2}$. As this is an arbitrary definition, we have tested that decreasing the threshold by orders of magnitude does not change the nature of our results. Therefore, we stick to this value for the sake of convinience in the numerical simulations. When $\tau$ is depicted versus $q$ in Figure~\ref{fig3}, a special value for $q$, $q^*$, for which $\tau$ is minimum is found.  This value $q^*$ represents thus an optimum mixture of the two imitation dynamics in order to achieve full coordination in the minimum time. The existance of a $q^*$ is robust to a change in the network topology as can be also seen in Figure~\ref{fig3}. The introduction of different topologies changes the particular value of $q^*$ but does not change the general scenario, as can be seen from the comparison of panels a and b (ER networks with different $\langle k  \rangle$) and panels c and d (scale-free networks with different $\beta$).

Another remarkable property of the dynamics of this system is the different way in which $n_A(t)$ decays when $q$ is above or below $q^*$. Figure~\ref{fig4} shows in detail the evolution of $n_A(t)$ for values of $q$ in each of these regimes.  For $q\lesssim q^*$ there is a power-law decay, which becomes exponential if instead $q\gtrsim q^*$. The identification of value of $q$ for which the functional form of the  $n_A(t)$ decay changes in nature is hard to obtain numerically but  within the uncertainty of our simulations that value seems consistent with $q^*$. 
%This leaves us with a description of the dynamics similar to one sketched in Fig.~\ref{fig3}c.

\begin{figure}
\begin{center}
\includegraphics[angle=0,width=10cm,clip]{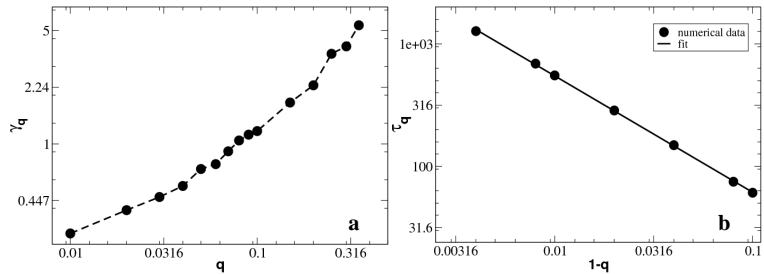}
\caption{Characterization of the functional decay of $n_A(t)$ in the two extremes of $q$. In both cases, the networks are Erd\"os-R\'enyi with $\langle k\rangle=9$ and a size of $N = 8000$. a: behavior of the power-law exponent of the active bond decay as a function of $q$ ($q<q*$).
b: behavior of the characteristic time in the exponential regime as a function of $1-q$. The fit gives $\tau_q\sim (1-q)^{-\xi}$ with $\xi\approx 0.95$.
\label{fig5}}
\end{center}
\end{figure}

The power-law decay of $n_A(t) \sim t^{-\gamma}$ for $q< q^*$ reserves us a final surprise. $\gamma$ changes with $q$, decreasing when $q\to 0$ (see Fig.~\ref{fig5}a). This and the analytical expressions found in the next calculations are reminiscent of the dynamics of glassy systems where a similar exponent decreases with the inverse of the temperature~\cite{bookglass}. 
As a final numerical result, Figure~\ref{fig5}b shows the apparent divergence of the characteristic time $\tau$ in the other limit, $q \to 1$, for an almost pure voter dynamics. As will be discussed next, this behavior, $\tau \sim 1/(1-q)$, can be understood with our analytical calculations.

\subsection*{Analytical approach}

The key for a correct understanding of how the dynamics works in our model is to evaluate the interplay of the mechanism of imitation at the local and global scales of the network. In order to shed light to the inner mechanisms of the dynamics, we make use of a very simple network model, where the two levels are clearly discernible. We consider a system composed of $M$ cliques (totally connected subgraphs), each one of $n+1$ nodes. Each node of a clique has $n$ connections internal to the clique and $C$ external ones. For the sake of simplicity, we assume that $C$ is equal for all the nodes and that it is a fraction $\alpha$ of the internal connections, $C= \alpha n$.

Now, let us focus on the voter dynamics. If $A_i$ is the number of agents playing one of the two choices, say choice $A$, in  clique $i$, and $\rho_i = A_i/(n+1)$ is the density of $A$ players, the variation in time of $\rho_i$ due to the voter contribution to the dynamics is
\begin{equation}
\label{evolVM}
\left[\frac{d\rho_i(t)}{dt}\right]_{VM}= \frac{1}{M}\, \frac{\alpha}{1+\alpha} \, (\rho-\rho_i) ,
\end{equation}
with $\alpha = n/C$ and where $\rho = \langle \rho_i \rangle$ is the average density of choice $A$ in the full network. Analogously, we can evaluate the contribution of unconditional imitation:
\begin{equation}
\label{rateCG_4}
\left[\frac{d\rho_i(t)}{dt}\right]_{UI}= \frac{1}{M} \, \left\{ \Theta\left[\rho_i - \frac{1}{2} + \alpha \, (\rho - \frac{1}{2})\right]-\rho_i(t)\right\} \ ,
\end{equation}
where $\Theta[\cdot]$ is the Heaviside function. Then, because each agent has a probability $q$ per time unit to evolve according to voter dynamics and $1-q$ to evolve through unconditional imitation, the general
evolution equation for $\rho_i$ is
\begin{equation}
\label{eveq}
\frac{d\rho_i(t)}{dt}=q \, \left[\frac{d\rho_i(t)}{dt}\right]_{VM} +(1-q) \, \left[\frac{d\rho_i(t)}{dt}\right]_{UI} .
\end{equation}

The final outcome of the game in the two extremes $q = 0$ or $q = 1$ can be obtained from Eqs.~(\ref{rateCG_4}) and (\ref{evolVM}), respectively. If $q=0$ (only unconditional imitation), the density $\rho_i$ tends to $\rho_i^{\infty}=0$ or $1$ for large values of $t$ and each clique will end up in a different frozen configuration
according to its initial conditions. On the other hand, if $q=1$ (only voter dynamics) it tends to $\rho_i=\rho$, that is, the system converges to an active disordered configuration conserving the overall density of choices. 
Finally, for intermediate $q$ values, if $0 < q < 1$, both conditions must hold to obtain a stationary state. This implies that the system must reach
full coordination.

In order to understand how the system converges to consensus close to the limits of $q$, it is convenient to sum up both members of equation~(\ref{eveq}) over the cliques, and
to divide by $M$. Since $(\sum_i\rho_i)/M=\rho$, we get a single expression for $\rho$
\begin{equation}
\label{aveveq}
\frac{d\rho(t)}{dt}=(1-q)\,\left\{\frac{1}{M}\sum_{i=1}^M\Theta\left[\rho_i(t)-\frac{1}{2}+\alpha\left(\rho(t)-\frac{1}{2}\right)\right]-\rho(t)\right\} \ .
\end{equation}
In the limit $q\to 1$, unconditional imitation is rare and we can assume that the voter dynamics is much faster. Taking into account that the system is going towards consensus ($\rho_i \to \rho$), it  can be shown that the term with the Heaviside function can be written as $\Theta(\rho-1/2)$
and then Equation~(\ref{aveveq}) becomes
\begin{equation}
\frac{d\rho(t)}{dt}\approx (1-q)\, \left(\Theta\left[\rho-\frac{1}{2}\right]-\rho\right) =
\left\{
\begin{array}{cc}
-(1-q)\rho & \mbox{if } \rho<  \frac{1}{2} \,, \\
(1-q)\left(1-\rho\right) & \mbox{if } \rho > \frac{1}{2}\, .
\end{array}
\right.
\end{equation}
If we focus on the active bond density $n_A(t)$ close to consensus where
$n_A\approx\rho(1-\rho)$, the previous equation implies that $ n_A\sim\exp[-(1-q)t]$,
from where one can see that the ordering is exponential and that the characteristic time scales as $\tau_q\sim\frac{1}{1-q}$. This behavior is numerically confirmed in Figure~\ref{fig5}b.

Calculations in the limit $q \to 0$ require a different approach. It is possible to justify the dependence of the exponent $\gamma$ on $q$ inserting a generic functional form
$\rho(t) \sim t^{-\gamma}$ into Equation~(\ref{aveveq}). For very small $q$ the dynamics is initially dominated by unconditional imitation, so that each clique reaches soon a
consensus state. In this way, the term with $\Theta[\cdot]$ in Equation~(\ref{aveveq}) counts only the cliques with choice $A$  and so it becomes close to the global variable
$\rho$ except for small fluctuations that vanish for $q$ going to zero. If we assume that the term with $\Theta[\cdot]$ can be substituted by $\rho + \delta$ and if, for instance,
the system is going towards consensus on the $B$ choice, then $d \rho/dt \approx d n_A/dt$ and  $|\delta|  \approx (1-q) \, \gamma \, t^{-(1+\gamma)}$. This relation implies
that $\gamma$ is a function of $q$, which vanishes for $q \rightarrow 0$ and diverges for a particular $q$ value that numerically we have identified with $q^*$.

\section*{Discussion}

In this work we have studied a game dynamics based on the interplay between strategic and random imitation. Our starting consideration is that in many socio-economic systems the process of imitation can be biased by social pressure as much as by strategic decisions. The model analyzed here is characterized by a parameter $q$ which relative weight of each of these  dynamics on the evolution of agents' choices. With
$q=1$, the system evolves by pure voter dynamics, which means that a random neighbor is imitated; on the other hand, for $q=0$ the agents take strategic decisions and copy the action of their best performing neighbor.

As an initial benchmark, we have considered regular topologies. We observed that any amount of mixing in the dynamic rules leads the system towards a final consensus where only one strategy survives. This occurs due to an opposite balance of forces: in complete graphs ({\it i.e} in the mean-field
approximation), where each agent is directly connected with everyone else and then has a perfect and complete information,
strategic decisions are needed to reach consensus, but on low-dimensional lattices, where agents have information
only about their closest proximity, a grade of random imitative dynamics is necessary for a population to reach a complete ordering (consensus for $q\neq0$).
In practice, the system needs special ``noises'' to avoid either the dynamic trap of the voter dynamics or the freezing onto local consensus incompatible with other regions of the network.

We then proceeded to study situations closer to real social networks, such as heterogeneous graphs. There, we found  a very interesting result: Pure dynamics, whether strategic or voter, leaves the system
disordered, but any amount of mixing of them allows to reach total consensus. Simplifying, 
we can state that if in low-dimensional
lattices noise is needed to reach consensus, and in mean field strategic incentive is necessary, on heterogeneous networks
both mechanisms are needed.
Moreover, there exist a perfect amount of mixing embodied by $q^*$ that leads to the fastest ordering of the system. The actual value of $q^*$ depends on the details of the particular network but it seems to exist always as long as the graphs is sparse and displays small-world effect. The dynamics of the system changes above and below $q^*$. In particular, it becomes extraordinarily slow for $q$ values close to zero with a behavior that reminds of glassy systems.

These results may contribute to the understanding of the social choice dynamics observed in real life situations. Indeed, a mechanism like the one we are considering here might be at work when people make choices about basically equivalent choices, such as phone providers, or computer brands, in which social interaction is relevant to the choice. If people were only strategic, we would observe complete market freezing in groups that chose different alternatives; if people were only imitative, they would be continuously changing their choice as strategic considerations played no role. Instead, in real life the market share of the different options evolves, generally tending to eliminate some choices, i.e., to ordering. If most choices were strategic (e.g., how many of my friends use a given phone company), as it is reasonable to expect, evolution would be very slow but noticeable, in agreement with real data. It is clear that other mechanisms may be at work here, but the one discussed here is a first attempt to explain the dynamics of choice in real social networks and we hope it can stimulate further work in this direction.

\section*{ACKNOWLEDGEMENTS}

D.\ V. and M.\ S.\ M. acknowlege support from the project FISICOS (FIS2007-60327) of the Spanish Ministry of Economy and Competitiviness (MINECO). J.\ J.\ R. receives funding also from the MINECO through the Ram\'on y Cajal program and through the project MODASS. A.\ S.\ acknowledges grants MOSAICO, PRODIEVO and Complexity-NET
RESINEE of MINECO and MODELICO-CM of 
Comunidad de Madrid.

\subsection*{Author Contributions} 

D.V., J.J.R., M.S.M. \& A.S. designed research, D.V. \& J.J.R. performed research, D.V., J.J.R., M.S.M. \& A.S analyzed the data and interpreted the results. All authors wrote, reviewed and approved the manuscript.

The authors declare no competing financial interests.

\end{document}